\documentclass{iau}

\usepackage{amsmath}
\usepackage{graphicx}
\usepackage{multirow}
\newcommand{\apj}{Astrophys.\ J.}

\newcommand{\apjs}{Astrophys.\ J. Supplement}
\newcommand{\aj}{Astron.\ J.}
\newcommand{\mnras}{Mon.\ Not.\ Royal Astron.\ Soc.}
\newcommand{\aap}{A \& A}

\newcommand{\prl}{Phys.\ Rev.\ Lett.}

\begin{document}

\lefttitle{P. Li}
\righttitle{Super cuspy dark matter halos due to baryon-driven contraction}

\jnlPage{1}{7}
\jnlDoiYr{2023}
\doival{10.1017/xxxxx}

\aopheadtitle{Proceedings of IAU Symposium 379}
\editors{P. Bonifacio,  M.-R. Cioni \& F. Hammer, eds.}

\title{Super cuspy dark matter halos of massive galaxies due to baryon-driven contraction}

\author{Pengfei Li}
\affiliation{Leibniz-Institute for Astrophysics, An der Sternwarte 16, 14482 Potsdam, Germany}

\begin{abstract}
The interplay between dark matter (DM) and baryons has long been ignored when building galaxies semi-empirically and observationally. Here I show that baryonic gravity leads to an adiabatic contraction of DM halos, which is most significant in massive galaxies. Ignoring this effect, the derived DM halos are not guaranteed in dynamical equilibrium. I present a new approach to deriving DM halos from rotation curves, which incorporates the adiabatic contraction. The compressed halos turn out super cuspy with respect to NFW halos, which require smaller baryonic contributions and less concentrated primordial halos. I also examine the semi-empirical approach to building galaxies, and find the adiabatic contraction can shift massive galaxies from the observed radial acceleration relation dramatically. Both approaches lead to super cuspy DM halos for massive galaxies, demonstrating the importance of the baryon-driven contraction, which has to be taken into account in order to make an apple-to-apple comparison with simulations.
\end{abstract}


\maketitle

\section{Introduction}

Galaxy formation in the standard $\Lambda$CDM model relies on a critical component, dark matter halos, which are invisible but set the very initial seeds for all the observed galaxies. To understand their structures, dark matter-only simulations have been employed, which suggest a Navarro-Frenk-White model \citep[NFW,][]{1996ApJ...462..563N}. To test these simulations, the properties of dark matter halos have been extracted from rotation curves for a variety of late-type galaxies. The comparison shows that while massive galaxies are consistent with simulations, dwarf galaxies present a cusp-core problem \citep[e.g. see][]{2002A&A...385..816D}: the NFW model predicts a halo that is too cuspy to fit observed rotation curves. Another way to test simulations is to build galaxies semi-empirically by assembling simulated dark matter halos and observed baryonic distributions into galaxies \citep[e.g. see][]{2016MNRAS.456L.127D,2017MNRAS.464.4160D}. One then tests whether the assembled galaxies are able to reproduce observed scaling relations established from dynamical studies of galaxies \citep[e.g.][]{2017MNRAS.471.1841N}. 

In both tests, the interplay between dark matter and baryons has generally been ignored, i.e. the two components are treated separately and independently, though they are closely coupled through mutual gravitational interaction. Baryonic gravity could possibly change the structure of dark matter halos, causing a net contraction. As a result, the built galaxies are not guaranteed in dynamical equilibrium. Therefore, whether the previous comparisons between observations and simulations are robust remains an open question to be tested.

\section{Significance of baryon-driven contraction}

To quantify the significance of the baryon-driven contraction, we can model the evolution of dark matter halos through an adiabatic process \citep{2006MNRAS.372.1869C}. \citet{1980ApJ...242.1232Y} introduced this approach by conserving three adiabatic actions and \citet{2005ApJ...634...70S} adopted it into the ``COMPRESS" code. Using this code, one can compute stabilized dark matter density profiles for any given baryonic distributions and dark matter halos.

The Spitzer Photometry \& Accurate Rotation Curves \citep[SPARC,][]{2016AJ....152..157L} database presents a representative sample of rotationally supported galaxies, from dwarfs to large spirals. Their NFW dark matter halos have been derived and presented in \citet{2020ApJS..247...31L}. To test the dynamical state of the best-fit halos, we compute the adiabatic changes using the ``COMPRESS" code \citep{2022A&A...665A.143L}. 

\begin{figure}[t]
  \centerline{\vbox to 6pc{\hbox to 10pc{}}}
  \includegraphics[scale=.35]{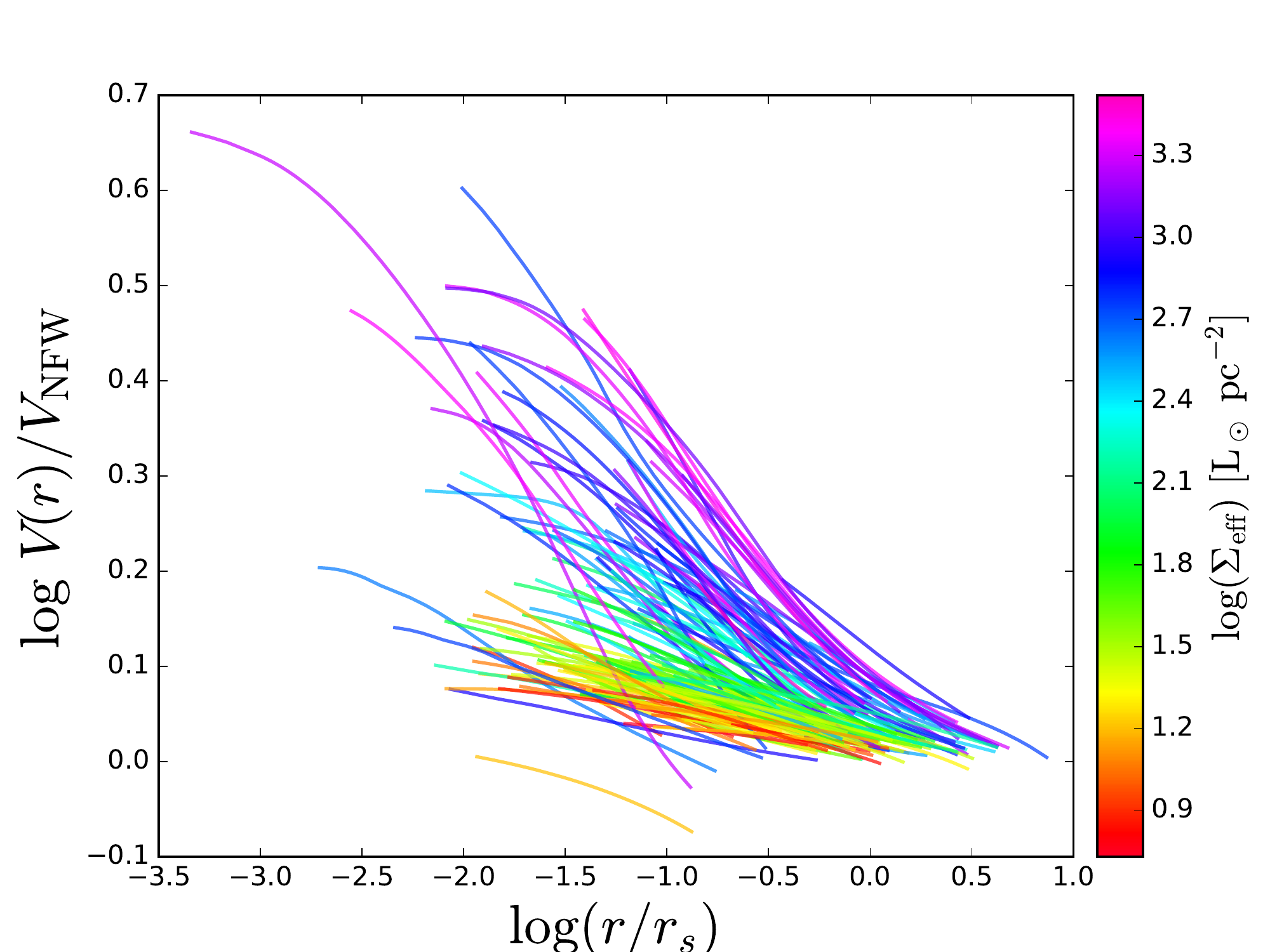}\includegraphics[scale=.35]{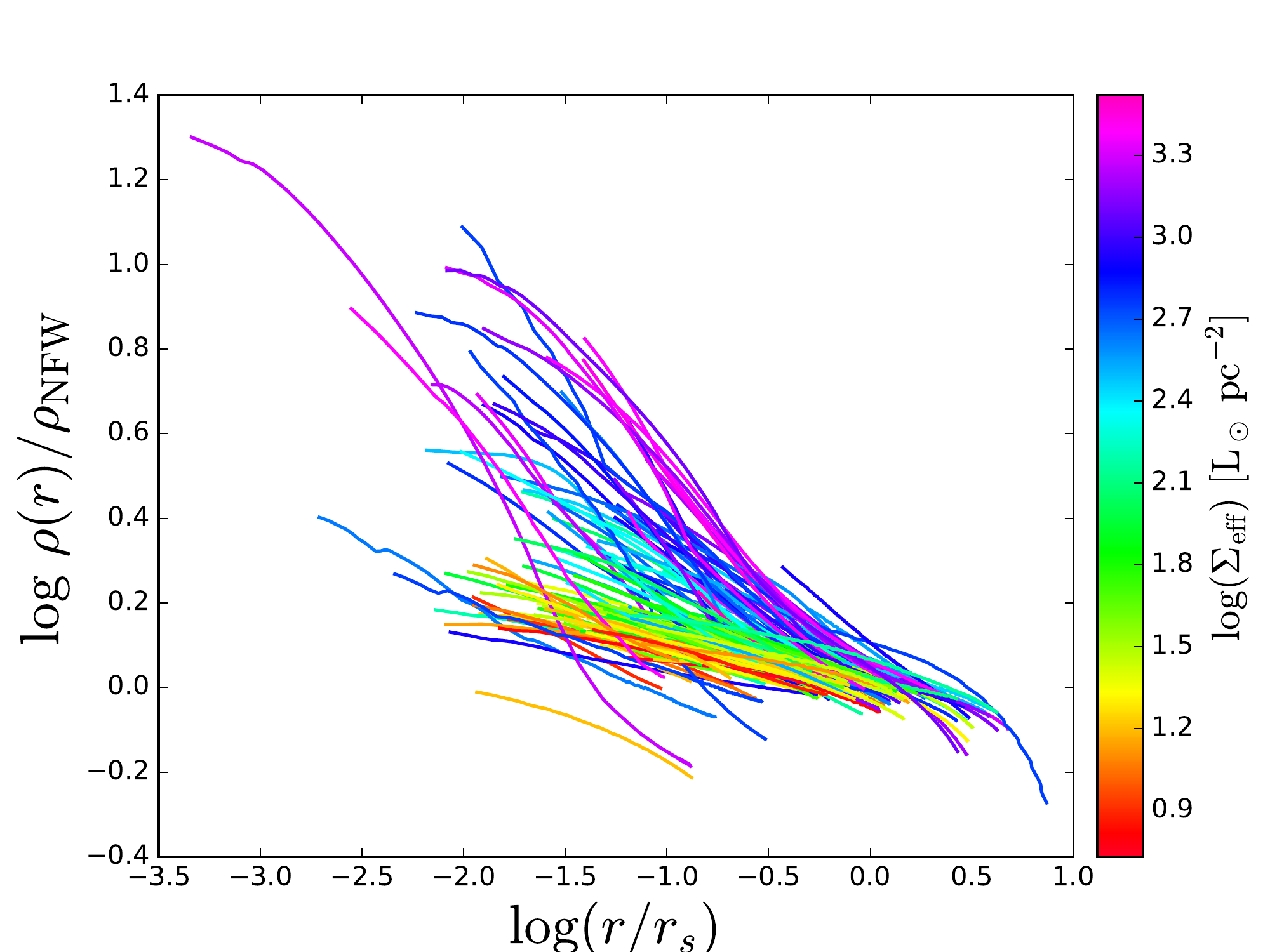}
  \caption{Left: Ratios of velocities contributed by compressed halos to primordial NFW halos; Right: density ratios between compressed halos and NFW halos. Lines are color-coded by their effective surface brightness. The initial NFW halos are from the best fits in \citet{2020ApJS..247...31L}.}
  \label{fig:compress}
\end{figure}

Figure \ref{fig:compress} shows the comparison between the best-fit NFW halos and the stabilized halos. Galaxies are color-coded by their effective surface brightness. For low-surface-brightness galaxies, their dark matter halos contract very little, so the best-fit NFW halos in \citet{2020ApJS..247...31L} are approximately in dynamical equilibrium. However, for galaxies with a surface brightness higher than 100 $L_\odot$ pc$^{-2}$, the dark matter halos experience strong adiabatic contraction. Their density profiles are significantly compressed by baryonic gravity. The inner density increases more significantly than the outer density, given the surface brightness is higher at small radii than at large radii. As a result, their contributions to rotation velocity increase dramatically. This breaks the obtained rotation curve fits, and it brings a dilemma for massive galaxies: the best-fit dark matter halos are not in dynamical equilibrium with embedded baryons, and the stabilized dark matter halos cannot describe observed rotation curves. 

\section{Deriving dark matter halos from rotation curves}

Since the traditional approach to deriving dark matter halos from rotation curves cannot guarantee the dynamical equilibrium of the whole system, it requires an improvement in the technique. This is made possible by combining evolving dark matter halos and fitting rotation curves \citep{2022A&A...665A.143L}. The new approach starts by assuming an initial dark matter halo and evolve it. One then calculates the stabilized halo using the ``COMPRESS" code and compare it with observed rotation curves. The process is iterated multiple times until a satisfactory fit is obtained. Since we fit stabilized halos to data, it guarantees the dynamical equilibrium for the coupled system. 

The initial halos can be parameterized by the NFW model, given simulations suggest that halos prior to galaxy formation are in the shape of NFW. The properties of the primordial NFW halos can be determined from the abundance matching relation by \citet{2013MNRAS.428.3121M} and halo mass-concentration relation by \citet{2014MNRAS.441.3359D}. However, there is no guarantee that these properties will be preserved after baryonic compression, even if they are imposed as priors.  

Though baryonic compression can significantly increase the inner density of dark mater halos, it is generally possible to achieve satisfactory rotation curve fits. This is because the strength of baryonic compression is quite sensitive to stellar mass-to-light ratios $\Upsilon_\star$. Reducing the value of $\Upsilon_\star$ leads to a lower surface mass density and thereby less significant adiabatic contraction. It also reduces the baryonic contribution to rotation curves, so that there is larger room for the contribution from dark matter halos. Since the net effect of baryonic compression is that halos become more concentrated, the concentrations for the initial halos are supposed to be smaller. In contrast, total halo mass is relatively less affected, as baryonic compression only changes the structure. \citet{2022A&A...665A.143L} indeed obtain good rotation curve fits for most of the SPARC galaxies, but the price is that both stellar mass-to-light ratios and initial halo concentrations are systematically turned down. 

\begin{figure}[t]
\centering
 \includegraphics[scale=.45]{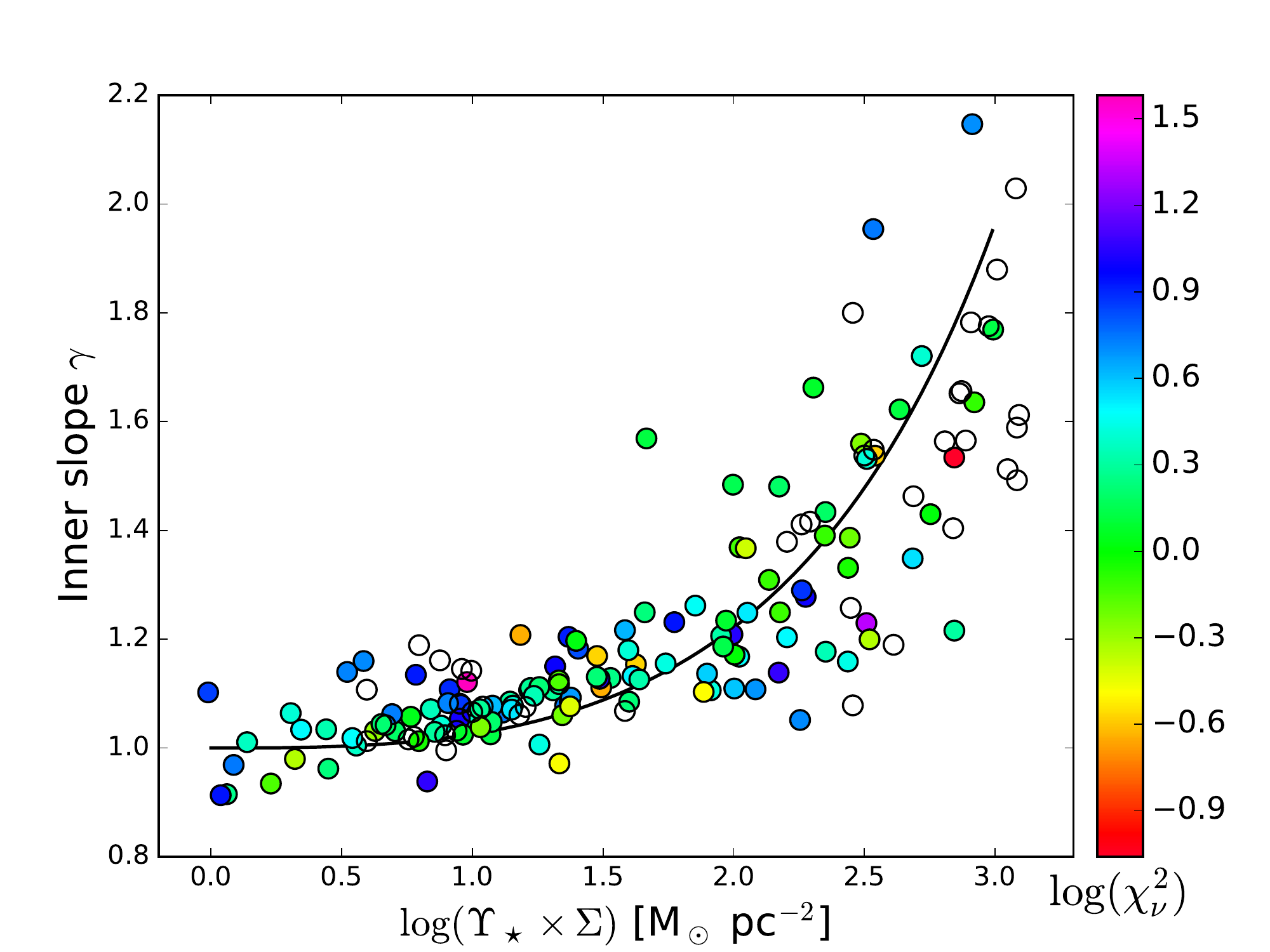}
  \caption{Best-fit values of the inner slope $\gamma$ as a function of surface mass density \citep{2022A&A...665A.143L}. Galaxies are color-coded by the fit quality. The solid line presents the selected exponential function that can roughly describe the mean relation.}
  \label{fig:gamma}
\end{figure}

In fact, the compressed halos are not even in NFW shape. To describe the stabilized halos, we can fit them with the generalized $(\alpha, \beta, \gamma)$ model \citep{1996MNRAS.278..488Z}. The fits suggest that the inner density slope $\gamma$ strongly correlates with surface mass density, as Figure \ref{fig:gamma} shows. The correlation can be roughly described by an exponential function,
\begin{equation}
    \gamma = \exp{\big(0.025(\log\Sigma_m)^3\big)},
\end{equation}
where $\Sigma_m=\Upsilon_\star\times\Sigma$ is the surface mass density. When the surface mass density is below 10 $M_\odot$ pc$^{-2}$, the inner slope is the same as NFW halos, suggesting that the baryonic compression is negligible. At higher surface mass density, the inner slope increases and the halos deviate from the NFW shape. For massive galaxies with $\Sigma_m>$ 100 $M_\odot$ pc$^{-2}$, the inner slope increases dramatically, forming super cuspy halos, which generally require relatively smaller stellar mass-to-light ratios to achieve satisfactory fits.

In the above calculation, I did not consider the effect of baryonic feedback. Feedback is believed to help convert cuspy halos to cores. The currently introduced mechanisms, such as supernova \citep[e.g. see][]{2014MNRAS.437..415D, 2014MNRAS.441.2986D}, stellar winds, star formation \citep[e.g.][]{2016MNRAS.459.2573R}, only work in dwarf galaxies. They are inefficient for massive galaxies, so the super cuspy halos we found for massive galaxies remain invariant. Other feedback mechanisms might work. Since the new approach considers the end product of the coupled system, it would be interesting to see whether feedback can keep counteracting baryonic compression all the time.

\section{Building galaxies semi-empirically}

The interplay between baryons and dark matter halos also plays an important role when building galaxies in a semi-empirical way. This approach sets up each component of galaxies separately and independently, such as stellar disks, gas disks, central bulges (if any) and dark matter halos. The detailed properties of each component can be set using results from observations (for baryonic components) and simulations (for dark matter halos). 

Many studies have been carried out using this approach \citep[e.g. see][]{2016MNRAS.456L.127D,2017MNRAS.464.4160D,2017MNRAS.471.1841N} to reproduce observed scaling relations in galaxies, such as the radial acceleration relation \citep[RAR,][]{2016PhRvL.117t1101M,2017ApJ...836..152L}, which is a tight correlation that holds in individual galaxies once the uncertainties on observational variables are taken into account \citep{2018A&A...615A...3L}. \citet{2017MNRAS.471.1841N} used this approach to setup galaxies and claimed that the RAR can be naturally explained by abundance matching relations. However, they did not consider the interplay between the baryonic components and the dark matter halos. So the galaxies they built may not be in dynamical equilibrium.

\begin{figure}[t]
  \centerline{\vbox to 6pc{\hbox to 10pc{}}}
  \includegraphics[scale=.35]{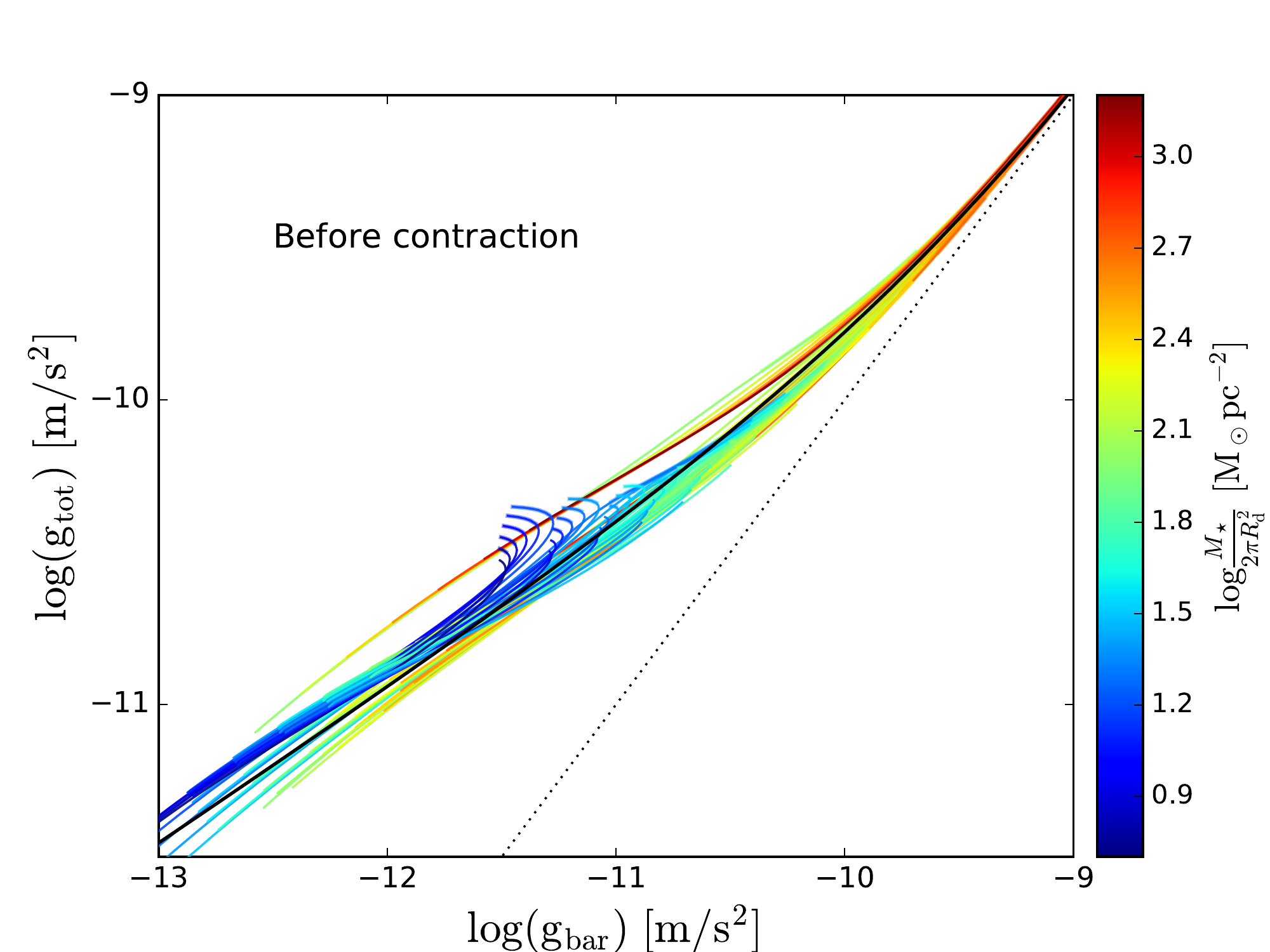}\includegraphics[scale=.35]{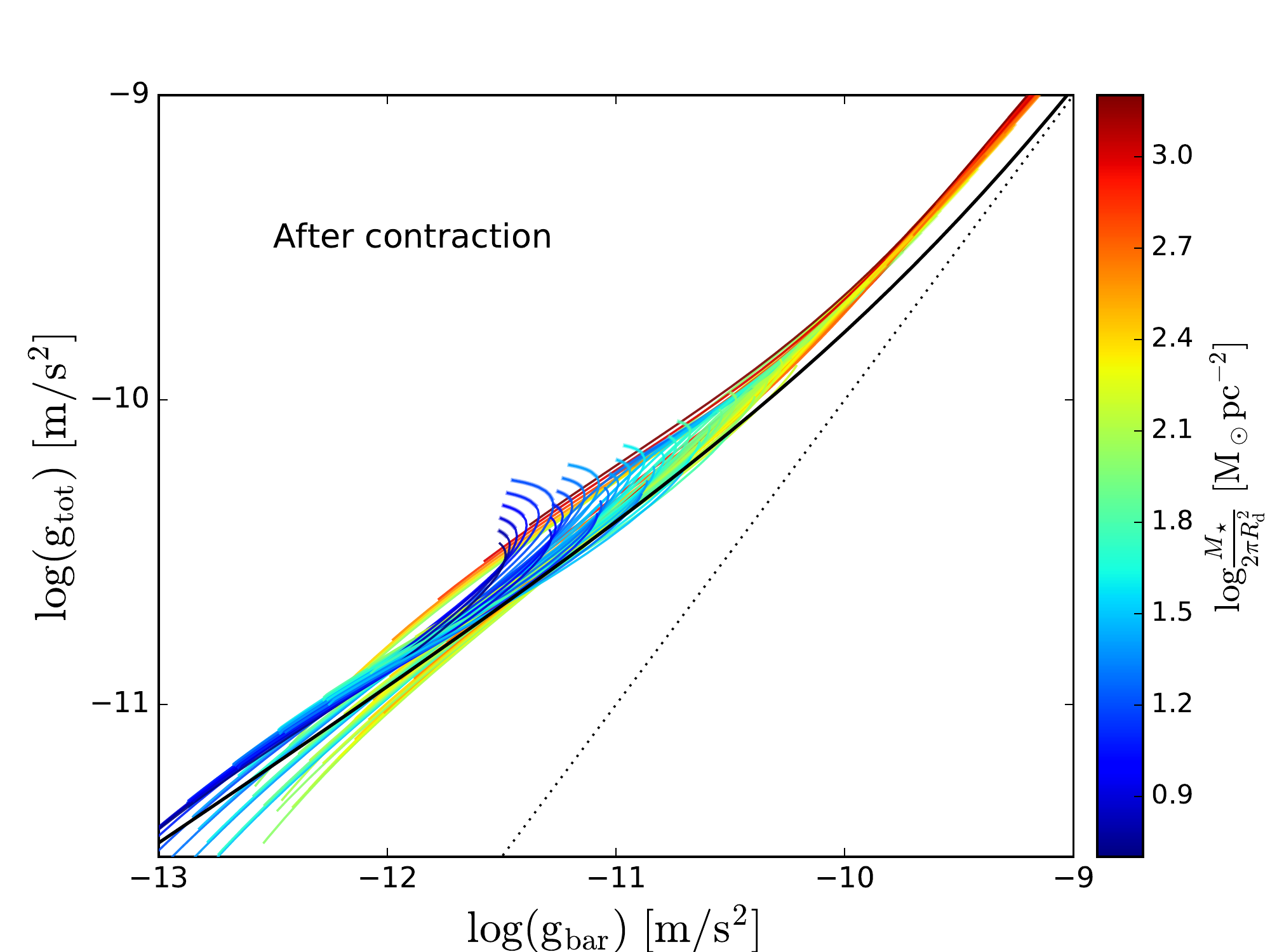}
  \caption{Radial acceleration relation from 80 galaxies built by setting up stellar disks, gas disks, central bulges (for massive galaxies) and NFW halos presented in \citet{2022ApJ...927..198L}. Galaxies are color-coded by their surface mass density. The left panel shows the results without incorporating the adiabatic contraction of dark matter halos, while the right panel shows the relation after adiabatic contraction.}
  \label{fig:RAR}
\end{figure}

To investigate how significant the interplay could affect the structure of dark matter halos, we build 80 model galaxies in \citet{2022ApJ...927..198L}. Stellar disks and gas disks are modeled with exponential profiles, and their properties (scale length and total mass) are chosen to match the SPARC galaxies \citep{2016AJ....152..157L}. For galaxies with $M_\star>10^9\ M_\odot$, a central bulge is included using the Hernquist profile \citep{1990ApJ...356..359H}. Similar to \citet{2017MNRAS.471.1841N}, dark matter halos are parameterized using the NFW model. One can employ abundance matching relation to determine the halo mass from their stellar mass. We find that the relation by \citet{2018AstL...44....8K} can generate the RAR that is most comparable to the observed one. Halo concentrations are determined using the mass-concentration relation from \citet{2014MNRAS.441.3359D}.

The left panel of Figure \ref{fig:RAR} shows that the 80 modeled galaxies nicely reproduce the observed RAR except for dwarf galaxies at small radii, which present some up-bended hooks. This is consistent with \citet{2017MNRAS.471.1841N}, but they cut off the inner parts, so their results do not present hooks. In contrast, we cut the modeled rotation curves based on the radial range of observed galaxies. The up-bended hooks are due to the cuspy NFW model, which is known to overestimate the inner density of dark matter halos. They can be removed if we model dark matter halos using cored halo profiles such as the DC14 model \citep{2014MNRAS.441.2986D}. Therefore, it is possible to build a sample of galaxies using the semi-empirical approach to reproduce the observed scaling relation by carefully choosing halo models and the abundance matching relation.

To examine the dynamical state of the model galaxies, we compute the stabilized halos by evolving the parameterized halos under the baryonic gravity. The right panel of Figure \ref{fig:RAR} shows the RAR after adiabatic contraction. While low-mass galaxies remain almost invariant, high-mass galaxies systematically shift upwards and significantly deviate from the observed RAR. This demonstrates that their dark matter halos experienced strong baryonic compression, and the inner mass density increases dramatically. As such, the total accelerations outweigh the baryonic ones at small radii, resulting in a net upward shift in the RAR panel. This suggests that the modeled NFW halos for massive galaxies turn to be super cuspy. Accommodating these super cuspy halos poses a challange for the CDM model. 

\section{Conclusion and outlook}

Dynamical tests demonstrate that the baryon-driven contraction is a quite significant effect for massive galaxies. To fairly compare observations with simulations, the adiabatic contraction has to be taken into account, and it leads to super cuspy dark matter halos that are difficult to be accommodated by observed rotation curves. 

There are two possible ways to get rid of these super cuspy halos. First, baryonic feedback is known to be able to reduce the inner density of dark matter halos, which is however most efficient in dwarf galaxies. For massive galaxies, one might need to introduce more powerful feedback processes. Second, the primordial halos may be less cuspy. Since we assume the primordial halos are NFW, which is already quite cuspy, it might be able to reproduce a normally cuspy halo if we start with core halos. Whether these two possibilities work out is worth further studies.

\begin{discussion}

\discuss{Hammer}{With this contracted halo for the Milky Way, what would be the ratio of the dark matter density to that of the baryons, at says, 6 kpc from the centre?}

\discuss{Li}{The ratio of the dark matter denisty to that of baryons at 6 kpc or the solar position should be higher than what we thought.}

\end{discussion}

\end{document}